\documentstyle[tighten,preprint,prd,aps]{revtex}

\begin{document}

\draft

\title{Gravitational Helioseismology? }
\author{Curt Cutler}
\address{Department of Physics, Pennsylvania State University}
\author{Lee Lindblom}
\address{Department of Physics, Montana State University}
\date{\today}
\maketitle

\begin{abstract} The magnitudes of the external gravitational
perturbations associated with the normal modes of the Sun are evaluated
to determine whether these solar oscillations could be observed with the
proposed Laser Interferometer Space Antenna (LISA), a network of satellites
designed to detect gravitational radiation.  The modes of relevance to
LISA---the $l=2$, low-order $p$, $f$ and $g$-modes---have not been
conclusively observed to date.
We find that the energy in these modes must be greater than
about $10^{30} \rm{ergs}$ in order to be observable above the LISA
detector noise.  These mode energies are larger than generally expected, but
are much smaller than the current observational upper limits.  LISA may
be confusion-limited at the relevant frequencies due to the galactic
background from short-period white dwarf binaries.  Present estimates of
the number of these binaries would require the solar modes to have
energies above about $10^{33} \rm{ergs}$ to be observable by LISA. 
\end{abstract}


\narrowtext
\section{Introduction}

The oscillation modes of the Sun are now routinely observed by
Doppler-shift measurements of the time-dependent velocity of the Sun's
surface~\cite{origobs,recentobs}.  The frequencies and amplitudes of
thousands of $p$-modes have now been accurately measured in this way.  A
number of detections of solar $g$-modes from Doppler-shift measurements
have also been reported, but the claimed detections all have low
signal-to-noise and they are not mutually consistent~\cite{palle}. 
It has recently been claimed (although, again, it is not universally
accepted) that certain periodic features observed in the solar wind
provide a second, completely independent, means of observing the solar
oscillations~\cite{windref}.  Some of the frequencies observed in the
solar wind correspond closely to already-observed solar $p$-modes, while
others are close to frequencies predicted for some solar $g$-modes. 

In this paper we investigate the possibility (also suggested
independently by Schutz~\cite{schutz} and by Gough~\cite{gough}) that
the solar oscillations might be observable using a third technique:
measuring the external gravitational perturbations associated with these
modes.  The proposed Laser Interferometer Space Antenna
(LISA)~\cite{lisa} is designed to detect the tiny gravitational
fluctuations caused by distant sources of gravitational radiation with
frequencies in the range $10^2$ to $10^6$ $\mu$Hz.  This detector would
also be capable of observing the fluctuations of the near-zone
gravitational field produced by the normal modes of the Sun, if the
amplitudes of the solar oscillations are sufficiently large.  This new
observational technique would be an interesting new probe of the
structure of the Sun because it would be most sensitive for detecting a
class of low-order modes that are presently unobserved using other
methods.  In this paper we calculate the magnitudes of the external
gravitational perturbations associated with the solar $p$, $f$, and
$g$-modes to estimate whether these oscillations could be observed with
LISA. 

\section{Solar Gravitational Perturbations}

We evaluate the coupling of the pulsation modes of the Sun to the
external gravitational field using a reasonably accurate model for the
structure of the Sun and its dynamics.  We use the equation of state of
a normal solar model (model 1 of Christensen-Dalsgaard, Proffitt, and
Thompson~\cite{cdpt}) to determine the equilibrium structure of the Sun. 
We treat the pulsations as small amplitude perturbations which satisfy
linear adiabatic evolution equations~\cite{Cox} with a realistic
dynamical adiabatic index~\cite{cdpt}.  We solve the equations for these
perturbations numerically to determine the magnitude of the external
gravitational perturbation that results from a mode of given amplitude. 
The gravitational perturbation in the exterior of the Sun has the form

\begin{equation}
\delta\Phi = \alpha_{28} \left({E\over 10^{28} {\rm ergs}}\right)^{1/2}
{GM_\odot\over R_\odot}\left({R_\odot\over r}\right)^{l+1}Y_{lm}e^{i\omega t},\label{1}
\end{equation}

\noindent where $r$ is the distance from the center of the Sun,
and $Y_{lm}$ is the standard spherical harmonic function.  The factor
$\alpha_{28}$ representss the magnitude of this external gravitational
perturbation for an oscillation mode normalized to have energy
$E=10^{28}$ergs.  The energy $E$ is defined as

\begin{equation}
E=\int \rho\,\, \delta v^*_a\delta v^a d^3x,\label{2}
\end{equation}

\noindent where $\rho$ is the mass density and $\delta v^a$ is the
(Eulerian) fluid velocity perturbation.  The values of $\alpha_{28}$
that we compute for a number of solar oscillation modes are presented in
Table~\ref{table1}.  The magnitude of the external gravitational
perturbation of a mode scales as $\sqrt{E}$ since $E$ is quadratic in
the perturbed fluid velocity.  The values of $\alpha_{28}$ given in
Table~\ref{table1} illustrate that the largest surface gravitational
perturbation occurs in the $l=2$ $g_{\,3}$-mode when modes excited to equal
energy levels are compared.  Also presented in Table~\ref{table1} are
the mode masses, $M_{\rm mode}$, defined as the ratio of the mode energy $E$
and the average surface velocity of the mode:

\begin{equation}
E= {M_{\rm mode}\over 4\pi}\int
\delta v^a\delta v^*_a \sin\theta d\theta d\phi.\label{2.5}
\end{equation}

\noindent The mode masses make it possible to convert the observable
surface velocities of modes into mode energies.  

In our numerical solution of the pulsation equations we use a simple,
but somewhat unrealistic, treatment of the surface of the sun.  We set
the (Lagrangian) perturbation in the pressure to zero at the point where
the pressure vanishes in our solar model.  For our equilibrium model we
use the realistic equation of state~\cite{cdpt} for densities above
$10^{-6}$gm~cm${}^{-3}$.  For densities below this value we smoothly
attach a polytropic ``atmosphere.'' The mode frequency $\omega$ and the
gravitational parameter $\alpha_{28}$ are rather insensitive to the
choice of density where this artificial atmosphere is attached.  However
the mode masses, $M_{\rm mode}$, which depend on the surface values of
the perturbed velocity, are rather more sensitive.  We estimate that the
errors in our computations of $\omega$ and $\alpha_{28}$ are about
0.01\%, while the errors in $M_{\rm mode}$ due to our simplified
treatment of the surface may be as much as 10\%. 

\section{LISA}

LISA is designed to detect the presence of gravitational radiation, or
other time dependent perturbations in the gravitational field, by
monitoring the precise distance between pairs of satellites.  Consider
two satellites which form one arm of the detector.  A near-zone
gravitational perturbation (such as that produced by the solar
oscillations) causes the distance between two satellites to oscillate
with the frequency $\omega$ of the gravitational perturbation.  The
amplitude $\delta L$ of the periodic displacement of the satellites is
determined by the standard tidal acceleration formula:

\begin{equation}
\omega^2\delta L = L \,n^a n^b \nabla_a\nabla_b \delta\Phi,\label{3}
\end{equation}

\noindent where $L$ is the average separation of the satellites.  The
unit vector $n^a$ that appears in Eq.\ (\ref{3}) gives the direction of
the line between the satellites.  To facilitate comparison with the
published LISA sensitivity curves, we define the dimensionless strain $h$
that would be measured by a detector consisting of two perpendicular 
interferrometer arms:

\begin{equation}
h  =  \omega^{-2}\,(n^a n^b - m^a m^b) \nabla_a\nabla_b \delta\Phi.
\label{hdef}
\end{equation}
 
\noindent where $n^a$ and $m^a$ are unit vectors pointing in the
directions of the two detector arms.  The right side of Eq.\
(\ref{hdef}) has a complicated dependence on the orientation of
the plane of the detector arms, the orientation of the
detector arms within this plane, the  position of the center-of-mass of 
the satellites, and the particular $l$ and $m$ of the mode.  For
simplicity we will use an average value for this quantity.  We
first average over all orientations of the two perpendicular arms, at a
fixed point in space, to obtain

\begin{equation}
 <\left| (n^a n^b - m^a m^b) \nabla_a\nabla_b \delta\Phi \right|^2> \ = \ 
{2 \over 5} \nabla_a\nabla_b \delta\Phi \nabla^a\nabla^b \delta\Phi^*.  
\label{have}
\end{equation}

\noindent We now average the right side of Eq.\ (\ref{have})
over all possible locations of the satellite around the Sun to find
the mean-square value of $h$,
 
\begin{equation}
<|h|^2>\  = \ {1\over {10\pi}}\int \nabla_a\nabla_b\delta\Phi
\nabla^a\nabla^b\delta\Phi^*\,\, \sin\theta d\theta d\phi.\label{4}
\end{equation}

\noindent Finally using Eq.\ (\ref{1}) we evaluate the angular integrals
on the right side of Eq.\ (\ref{4}) to find the average strain $h$ that
would be sensed by the LISA detector

\begin{eqnarray}
h = &&
\alpha_{28} \sqrt{(l+1)(l+2)(2l+1)(2l+3)} \nonumber\\
&&\times \left({1\over 10 \pi}\right)^{1/2}
\,\left({E\over 10^{28} {\rm ergs}}\right)^{1/2}
{GM_\odot\over \omega^2 R_\odot^3}
\left({R_\odot\over r}\right)^{l+3}.\label{6}
\end{eqnarray}

The modes of the Sun with the same $n$ and $l$ but different $m$ 
have frequencies that are slightly split by the Sun's rotation. 
In one year of integration, LISA will be able to
distinguish frequencies to within about $\Delta f \approx 3 \times
10^{-2}\mu$Hz (for signal-to-noise of order unity).  This $\Delta f$ is
roughly an order of magnitude smaller than the size of the splitting
due to the Sun's
rotation.  Thus modes with different $m$ are effectively non-degenerate,
and the energy $E$ that appears in Eq.\ (\ref{6}) must be that of a
single mode of given $l$ and $m$.

\section{Signal Strenghs and Noise Levels}

To determine whether the oscillation modes of the Sun could be detected
by LISA we must determine whether the expected gravitational signals
exceed the expected noise in the detector and any ``background''
gravitational wave signals.  We now give estimates for the parameters
that determine the signal strengths and the noise levels relevant to
this problem. 

As we have seen, Eq.\ (\ref{6}), the magnitude of the signal in the LISA
detector is determined by the energy contained in the modes of the Sun. 
The modes which are most likely to excite the LISA detector have not
been observed to date.  This makes the prospects of gravitational
observations of these modes very interesting, but it also makes it very
difficult to make reliable predictions of the energy contined in these
modes.  The observed low-$l$ p-modes have a maximum energy/mode of
roughly $10^{28} $ ergs for $f$ near $3000 \,\mu$Hz, with the
energy/mode decreasing to roughly $10^{27}$ ergs at $f$ near $1000
\,\mu$Hz \cite{libbrecht}.  A simple extrapolation of this curve would
suggest even lower energies for the modes relevant to LISA ($100$ to
$700 \,\mu$Hz), but such an extrapolation might well be naive. 
Goldreich and Murray have argued that the decrease in p-mode energy (at
fixed $l$) from $3000$ to $1000\,\mu$Hz is due to the scattering of
energy from higher-$n$ to lower-$n$ p-modes of the same frequency
\cite{goldreich_murray}.  But the p-modes relevant to LISA are the
lowest-$n$ p-modes, so they would not be damped by this mechanism. 

At present only upper limits for the energies of these low-frequency
modes are known.  This is mostly because, for a given $l$ and mode
energy, the detection of modes by Doppler shift measurements becomes
much more difficult as one goes to lower frequency modes.  This
difficulty is due to the fact that the the surface velocity (for fixed
energy) decreases roughly like $ f^{3.5} $ for $p$-modes and $f^{2}$ for
$g$-modes, while instrumental and background solar velocity noise both
increase at lower frequencies \cite{palle}.  The present observational
upper limit on the surface velocities of these modes is about
4~cm~s${}^{-1}$~\cite{palle}.  Using the mode masses computed in
Table~\ref{table1}, this translates into limits on the energies,
$E_{\max}$, which are also listed in Table~\ref{table1} for each mode. 
It has been suggested~\cite{palle} that the low signal-to-noise
``detections'' of solar g-modes are an indication that the $g$-mode
energies are in fact just below these maximum values.  These energies
are rather large, and many Solar modes would be well above the LISA
noise levels if their energies are close to these values. 

Kumar, Quataert, and Bahcall~\cite{kqb} have recently made predictions
of the energies contained in the low-frequency solar g-modes.  They
consider the turbulent convection excitation mechanism used by
Goldreich, Murray, and Kumar~\cite{gmk} to explain the excitations of
the observed $p$-modes.  Applying this mechanism to the low-$l$, low-$n$
$g$-modes, they find that turbulent convection would excite surface
velocities of about $10^{-2}$cm~s${}^{-1}$ in these modes, assuming the
dissipation time for the modes to be about $10^6$ years.  These surface
velocites correspond (using our computed mode masses) to energies
between $2\times10^{27}$ and $2\times10^{29}$ergs for the $g$-modes
listed in Table~\ref{table1}.  Thus, a typical value for the energy
predicted by this model is about $10^{28}$ergs.  Similar energies are
predicted for the low-$l$, low-$n$ p-modes~\cite{kumar_private}. 

For frequencies below $1000\mu$Hz the anticipated noise in the LISA
detector~\cite{Thorne} would prevent the detection of periodic signals
with amplitudes less than

\begin{equation}
h_{sp}=3 \times 10^{-22}\left({f\over 100\mu{\rm Hz}}\right)^{-2.3},\label{7}
\end{equation}

\noindent where $f=\omega/2\pi$ is the frequency of the
signal~\cite{rms}.  It is also possible that other gravitational wave
sources might produce a background ``noise'' from which the Solar
oscillation signals could not easily be distinguished.  Hils, Bender,
and Webbink~\cite{hils} have estimated that there are about $3 \times
10^6$ short-period white-dwarf binaries in our galaxy, producing a
gravitational wave background of about $30\, h_{sp}$ in the relevant
frequency band.  At the time these estimates were made there were no
known examples of white dwarf binaries with periods less that one day. 
Recently, however, two white dwarf binaries with periods of $3.47$ and
approximately $4$~hours resectively were discovered by
Marsh~\cite{marsh} and Marsh, Dhillon, and Duck~\cite {marshetal}. 
Given these new observations, it should be possible to make considerably
better estimates of the gravitational wave background from binaries, but
to our knowledge, such estimates have not yet been completed.

\section{Results}

Table~\ref{table1} lists the values of the strain amplitude $h$, Eq.\
(\ref{6}), for a number of modes of the Sun, assuming that each mode has
an energy of $10^{28}$ergs and that the detector is located at a
distance $r=1\,$AU from the Sun.  Only the $l=2$ modes are considered,
since the modes with successively higher $l$ values have gravitational
signals which are reduced below their $l=2$ counterparts by additional
factors of $R_\odot/r\approx 0.005$.  The modes in Table I are listed in
order of increasing $E_{\rm min}$, the mode energy for which
$h=3h_{sp}$.  This is the minimum energy needed in a mode so that the
mode would be observable above the detector noise with $S/N=3$. 
Included in Table~\ref{table1} is every $l=2$ mode with $E_{\rm min}\leq
10^{32}$ergs.  We see that mode energies of at least $10^{30}$ergs are
required for any of them to be observable by LISA.  This is a factor of
(up to) $10^3$ smaller than the upper limits, but a factor of $10^2$
larger than the energies estimated on theoretical grounds by Kumar,
Quataert, and Bahcall~\cite{kqb}.  Given these energies, the
corresponding surface velocities of these modes would be about
$10^{-1}$cm~s${}^{-1}$, which is roughly the detection threshold for the
recently-launched SOHO satellite~\cite{gabriel}.  This indiciates that
any modes that could be observed by LISA would already have been
observed by SOHO.  Nevertheless, LISA could still provide some unique
information since it measures the gravitational amplitude of the mode
which depends on the density perturbations throughout the interior of
the Sun. 

  If the background radiation due to short-period white dwarf binaries
is comparable to the values predicted by Hils, Bender, and Webbink
\cite{hils}, then at $f \approx 300\,\,\mu$Hz the gravitational wave
background is a factor of about $30$ larger than LISA's detector noise. 
Hence mode energies about $10^3$ times greater than the $E_{\min}$'s
listed in Table~\ref{table1} would be required for solar oscillations to
rise above this background.  Such energies would be close to the present
upper limits for these modes.  Therefore, if the gravitational wave
background is close to the strength predicted by Hils, Bender and
Webbink~\cite{hils}, then the mode energies would have to be very close
to their maximum allowed values for LISA to contribute at all to
helioseismology.  LISA's one-year orbit around the Sun would modulate
the solar oscillation signal in a way that differs from the one-year
modulation of the background gravitational wave signal.  It is
conceivable that this difference could be used to help distinguish solar
oscillations from the gravitational wave background, but we have not
investigated this possibility in any detail.

\acknowledgments We thank Peter Bender, Lars Bildsten, Steven Detweiler,
Pawan Kumar, James Ipser, and Norman Murray for helpful discussions. 
C.C.  was supported by NSF grant PHY-9507686 and an Alfred P.  Sloan
Fellowship.  L.L.  was supported by NSF grant PHY-9019753 and NASA grant
NAGW 4866.

\mediumtext
\begin{table}
\caption{Gravitational Parameters of Solar Oscillations.}
\label{table1}
\begin{tabular}{ccccccc}
Mode &$f$ ($\mu$Hz) &$\alpha_{28}$ &$M_{\rm mode}$ (gm) 
&$h$ &$E_{\min}$(erg) &$E_{\max}$(erg)\\
\tableline
$g_{\,3}$ &$220.4$ &$7.17\times 10^{-12}$  &$8.3\times 10^{31}$
&$1.18\times 10^{-23}$  
&$1.55\times 10^{30}$ &$1.3\times10^{33}$ \\

$g_{\,2}$ &$254.0$ &$6.07\times 10^{-12}$  &$4.4\times 10^{31}$
&$7.50\times 10^{-24}$  
&$1.97\times 10^{30}$  &$7.0\times10^{32}$\\

$p_{\,1}$ &$381.6$ &$5.31\times 10^{-12}$  &$4.3\times 10^{30}$
&$2.91\times 10^{-24}$  
&$2.02\times 10^{30}$  &$6.9\times10^{31}$\\

$g_{\,4}$ &$192.2$ &$5.96\times 10^{-12}$  &$1.9\times 10^{32}$
&$1.29\times 10^{-23}$  
&$2.42\times 10^{30}$  &$3.0\times10^{33}$\\

$g_{\,5}$ &$168.8$ &$4.72\times 10^{-12}$  &$3.5\times 10^{32}$
&$1.32\times 10^{-23}$  
&$4.19\times 10^{30}$  &$5.6\times10^{33}$\\

$g_{\,1}$ &$293.6$ &$2.81\times 10^{-12}$  &$2.3\times 10^{31}$
&$2.60\times 10^{-24}$  
&$8.44\times 10^{30}$  &$3.7\times10^{32}$\\

$f$ &$350.9$ &$2.31\times 10^{-12}$  &$1.2\times 10^{31}$
&$1.50\times 10^{-24}$  
&$1.12\times 10^{31}$  &$1.9\times10^{32}$\\

$p_{\,2}$ &$514.4$ &$2.02\times 10^{-12}$  &$5.5\times 10^{29}$
&$6.07\times 10^{-25}$  
&$1.17\times 10^{31}$  &$8.8\times10^{30}$\\

$g_{\,6}$ &$134.0$ &$2.80\times 10^{-12}$  &$8.2\times 10^{32}$
&$1.24\times 10^{-23}$  
&$1.37\times 10^{31}$  &$1.3\times10^{34}$\\

$p_{\,3}$ &$663.6$ &$8.60\times 10^{-13}$  &$1.4\times 10^{29}$
&$1.56\times 10^{-25}$  
&$5.53\times 10^{31}$  &$2.2\times10^{30}$\\

$g_{\,7}$ &$101.2$ &$1.42\times 10^{-12}$  &$1.6\times 10^{33}$
&$1.10\times 10^{-23}$  
&$6.30\times 10^{31}$  &$2.6\times10^{34}$\\
\end{tabular}

\end{table}


\begin{references}
\bibitem{rms}  The function $h_{sp}$ is defined so that periodic source 
producing a signal at this level for a one-year integration time could be
detected at the $1\,\sigma$ level.
\end{references}
\end{document}